\documentclass[10pt]{article}

\usepackage{amsmath}
\usepackage{amssymb}
\usepackage{rotating}
\usepackage{graphicx}
\usepackage{epstopdf}
\usepackage{subfigure}
\usepackage{cite}
\usepackage{multirow}
\usepackage{color} 

\usepackage[labelfont=bf,labelsep=period,justification=raggedright]{caption}

\makeatletter
\renewcommand{\@biblabel}[1]{\quad#1.}
\makeatother

\date{}

\begin{document}

\begin{flushleft}
{\Large
\textbf{How the Taxonomy of Products Drives the Economic Development of Countries}
}
\\
Andrea Zaccaria$^{1,\ast}$, 
Matthieu Cristelli$^{1}$, 
Andrea Tacchella$^{1,2}$, and
Luciano Pietronero$^{1,2,3}$
\\
\bf{1} ISC-CNR, Via dei Taurini 19, 00185, Roma, Italy.
\\
\bf{2} Dipartimento di Fisica, Sapienza Universit\`a di Roma, P.le Aldo Moro 2, 00185, Roma, Italy.
\\
\bf{3} LIMS, London Institute for Mathematical Sciences, 35a South Street Mayfair, London, UK
\\
$\ast$ E-mail: andrea.zaccaria@roma1.infn.it
\end{flushleft}

\section*{Abstract}
We introduce an algorithm able to reconstruct the relevant network structure on which the time evolution of country-product bipartite networks takes place. The significant links are obtained by selecting the largest values of the projected matrix. We first perform a number of tests of this filtering procedure on synthetic cases and a toy model. Then we analyze the bipartite network constituted by countries and exported products, using two databases for a total of almost 50 years. It is then possible to build a hierarchically directed network, in which the taxonomy of products emerges in a natural way. We study the influence of the structure of this taxonomy network on  countries' development; in particular, guided by an example taken from the industrialization of South Korea, we link the structure of the taxonomy network to the empirical temporal connections between product activations, finding that the most relevant edges for countries' development are the ones suggested by our network. These results 
suggest paths in the product space which are easier to 
achieve, and so can drive countries' policies in the industrialization process. 
\section*{Introduction}
The study of how countries develop has a central role in Economics and major consequences in political, industrial and financial analyses and evaluations. Historically, a number of approaches have been applied to this problem. According to the model introduced by Heckscher and Ohlin \cite{heckscher1991heckscher}, which is based on Ricardo's comparative advantage \cite{ricardo1891principles}, the possible pattern of progress of a country is a direct consequence of its endowments, namely the presence of productive factors such as land, labor and capital. This approach has been challenged by Leontief \cite{leontief1956factor,leontief1954domestic}, who found a striking empirical counterexample, now known as Leontief's paradox (but see also \cite{leamer1980leontief} for a contrary view) and, again on an empirical basis, by Bowen et al. \cite{bowen1987multicountry}. Another approach has been proposed by Aghion and Howitt \cite{aghion1992model}, whose model is inspired on the concept of creative destruction, 
originally introduced by Schumpeter \cite{schumpeter1934theory}, which focuses on endogenous factors such as technology. This perspective has originated from the seminal paper by Romer \cite{romer1990endogenous}. As pointed out, for example, by Hausmann and Rodrick \cite{hausmann2003economic}, these views can be summarized in the assumption that the basic needs for a sustained growth are tradable technology and good local institutions. Hausmann and Rodrick in the same paper give two examples, the growth of some Asian countries and the recession of the Latin American ones, in which the opposite was true. For example, South Korea and Taiwan experienced an impressive growth even if they retained high levels of protection, while Latin American countries performed better in the decades 1950-1980, with poorer institutions, than in the 90's, when their governments adopted the long awaited structural reforms. Lall \cite{lall2000technological} suggested a third line of reasoning, which he calls the ``capabilities 
approach". A capability of a country can be, in its wider sense, anything which makes the country able to produce a given product, from infrastructures to efficient scholar and administrative institutions, from a mild fiscal policy to demography issues. According to Lall, the crucial point is not the simple knowledge of a technology, but the ability to exploit its potential, that is to be able to use it efficiently given the intrinsic properties of the specific country. As a consequence, each country has to find its own path towards development, focusing on its learning system in order to add capabilities to the ones it already owns. This line of reasoning, in which each country has to learn first what one is good at producing and then which technology can be best adapted to its case, has been modeled in a general equilibrium framework in \cite{hausmann2003economic}. By contrast, our approach is closer to the concept of \textit{adjacent possible}, introduced by Kauffman \cite{kauffman2002investigations} and 
originally applied to biological systems \cite{stuart1993origins}. Finally, we mention the evolutionary approach \cite{nelson2009evolutionary}, in which the optimizing role of the market is substituted by a 
natural selection process which assures a ceaseless change, in general, of any economic process. This ideas gave rise to new fields of research, such as the ones regarding innovation \cite{fagerberg2009innovation}.\\
The time series of exports give a fundamental insight to understand countries' development, and can help define an empirical framework to assess the validity of theoretical paradigms. If we suppose that products are defined by means of the set of capabilities which are needed for a country to be able to produce it, the presence or the absence of a product in a country's export basket represent a hint on the capability basket of the country itself. In particular, one can build a network of products in which two products are connected if they share some capabilities; in practice, if many countries produce the same couple of products. In this way, one can avoid to study the capabilities structure of countries, which is, at best, very hard to represent or even define from a quantitative point of view. This network, called the Product Space, has been introduced and studied by Hidalgo et al. \cite{productspace} (see also \cite{hidalgo2009dynamics}, and \cite{caldarelli2012network} for a different approach). In the 
present work we want to build a different network of products, in which two nodes are connected by a directed link which represents the causality relationship between them. For example, two products $a$ and $b$ will be connected not if they are similar, but if one of them, say $a$, makes more probable that $b$ will be produced in the future. In this case, the directed link will go from $a$ to $b$. We propose an algorithm which suitably filters the information contained in the empirical export data and permits to build a hierarchical network whose nodes are products and the directed links are given by the necessity relationship between products. In this structure, in which the link between capabilities and development emerges in a clear way, the number of edges is reduced with respect to the almost fully connected network which can be obtained by a simple projection of the bipartite country-product network; namely, we reduce the number of links from about $N^2$ to order $N$ by selecting the most informative 
ones from the point of view of economic progress. We find that while the development of many countries is mostly driven by their initial conditions, most of them walk through recurrent paths, suggesting the presence of mandatory steps in the industrial progress of nations.\\

\section*{Materials and Methods}

\subsection*{Data description}

To build our network we use two databases, both reporting the import-export flows between countries. The first is collected by the United Nations  and processed by Feenstra et al. \cite{feenstra2005world} and concerns the years from 1963 to 2000, while the second is collected by the United Nations and processed by BACI \cite{abbracci} and covers the years from 1995 to 2010. The number of countries ranges from 134 to 151. After a detailed cleaning process, whose aim is to remove clear errors and inconsistencies, we build a matrix $M_{cp}$ whose elements are equal to 1 if country $c$ exports product $p$ and zero otherwise. These values are assigned using a threshold on the Revealed Comparative Advantage, as introduced by Balassa \cite{balassa1965trade}. This way of organizing this kind of databases has already been proved to be useful in the quantification of the growth potential of countries \cite{HH,newmetrics,plosnm}. Clearly, each year has a defined import-export structure, so the resulting matrices are 
different; moreover, the two databases have a different number of products, 
because they are categorized with different classifications. As a result of the cleaning process the number of products in each one of the two databases is kept constant through the years. So we will build and analyze two networks: the first one, referring to the years 1995-2010, contains 1131 products classified in HS2007; the second one, spanning from 1963 to 2000, has a lower number of products (538), classified in SITC rev.4 and permits an analysis of the development of countries on a longer time horizon, spanning several economic cycles.

\subsection*{Taxonomy and Proximity}

\begin{figure}
\centering
\subfigure[Taxonomy Network]{
\includegraphics{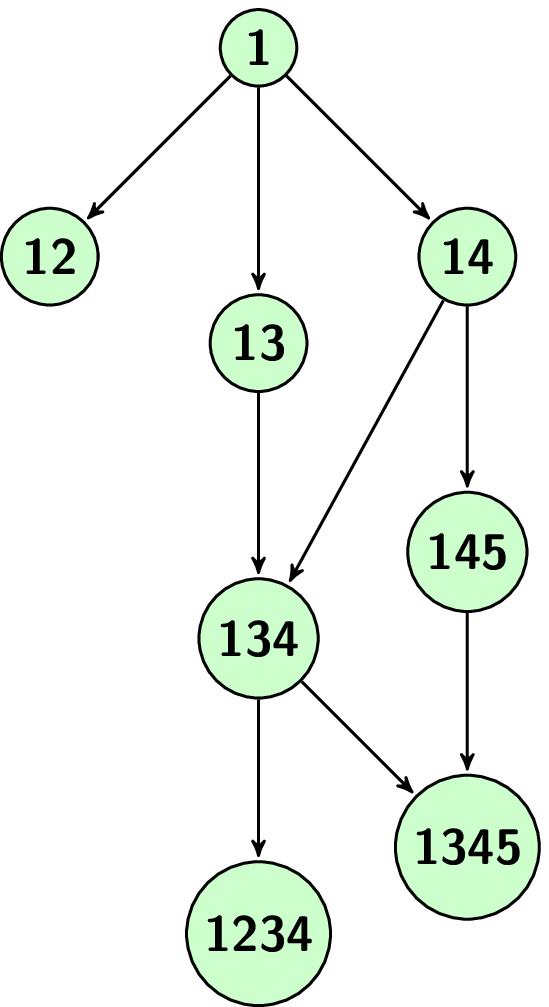}
}
 \hspace{10mm}
\subfigure[Proximity Network]{
\includegraphics{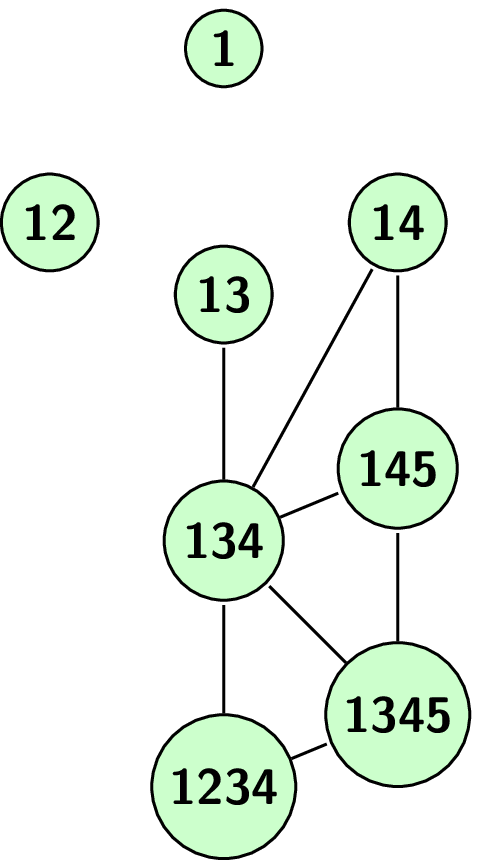}
}
\caption{Two different ways to connect the same products, which are characterized by the capabilities needed to produce them. On the left, we consider a hierarchical relationship; on the right, we join similar products.} \label{fhsdsljfjks}
\end{figure}

As we anticipated above, differently from the approach described by Hidalgo et al.\cite{productspace}, we want to build a hierarchically ordered network, whose structure is inferred from the $M_{cp}$ matrix. The idea can be easily understood by means of the concept of capability \cite{lall2000technological,HH}. Let us define the products in terms of the capabilities which are needed to conceive and produce them. For example, the capability 1 corresponds to a basic product. A country equipped with a second capability, 2, can export the ``12'' product. Capabilities 1,2 and 3 could simply not lead to a product, while ``134'' can be a product, and so on. A hierarchy naturally arises, in which some products are mandatory intermediate steps to be able to produce more complex technologies, and the \textit{sons} are connected to the \textit{father} by a directed edge. In Fig.\ref{fhsdsljfjks}(a) we show a possible example of this kind of structure, that we call a \textit{taxonomy} network. On the other hand, Fig.\
ref{fhsdsljfjks}(b) shows an example of a \textit{proximity} network, in which the same products are connected if they share a fraction, in this particular example more than one half, of their composing capabilities. In this case, one will have an undirected network, because the products are connected if they are similar, and so they are at the same level.\\ We want to stress that, when one takes into consideration real data, we expect that a country will likely move from basic products to more complex ones when it develops new capabilities. Thus, the time evolution of the technological progress should be closer to a taxonomy than to a proximity network.\\

\subsection*{Algorithm description}

Let us define the \textit{diversification} $d_c$ as the number of products exported by the country $c$, as measured by the Revealed Comparative Advantage:
\begin{equation}
 d_c=\sum_{p} M_{cp}
\end{equation} 
and the ubiquity $u_p$ as the number of countries which export the product $p$:
\begin{equation}
 u_p=\sum_{c} M_{cp}.
\end{equation}
In order to obtain a product-product matrix we project $M_{cp}$:
\begin{equation}
 B_{pp'}=\frac{1}{\max(u_p,u_{p'})}\sum_{c} \frac{M_{cp} M_{cp'}}{\sqrt{d_c}}
 \label{hdksajhdk}
\end{equation}
this way of normalize the projection is similar to the one introduced by Zhou et al.\cite{zhou2007bipartite}. The $\sqrt{d_c}$ factor takes into account the different contribution given by countries of different diversifications, by dividing the corresponding terms by the expected values in a random binomial case. Nevertheless, since the exponents of ubiquity and diversification are somehow arbitrary, we have checked a posteriori their goodness by means of the toy model and the sample matrices discussed in the next section. Moreover, in order to obtain the adjacency matrix of a network with number of edges of the same order of magnitude of the number of products we select only the maximum entry of each row, excluding the diagonal elements. In other words, for each product $p$ we look for the product $p'\neq p$ which maximizes the normalized probability $B_{pp'}$ to be exported in a pair. Possible degeneracies are removed by looking at which product contributes the most with respect to its column; in other 
words, we pick the product whose column has the smallest elements. As we will show in the following, this filtering procedure is able not only to discard redundant and noisy information but also to define a set of 
preferred patterns for industrialization and development policies.\\
We point out that this procedure, in principle, could be applied using only the data which refers to one year, while we actually have 38 different matrices in the first dataset and 16 in the second dataset. While it would be natural to apply this algorithm
for each matrix of every year, we preferred to aggregate them in a single matrix with the same columns (the 538 or 1131 exported products) and, as rows, all countries, including repetitions due to different years. In this way, most of the fluctuations are averaged out.\\
In the following section we will describe the properties of the Taxonomy network we obtained by applying our algorithm on the complete $M_{cp}$ matrix.\\
For the sake of completeness we mention that, using our notation, the Product Space introduced by Hidalgo et al. is based on the proximity $\phi_{pp'}$ between the products $p$ and $p'$, which is defined as\cite{hidalgo2009dynamics}:
\begin{equation}
 \phi_{pp'}=\min\bigg(\frac{\sum_{c} M_{cp} M_{cp'}}{u_p},\frac{\sum_{c} M_{cp} M_{cp'}}{u_{p'}}\bigg). 
\end{equation}
This expression is quite similar to Eq.\ref{hdksajhdk} and, when used without any further filtering process, leads to an almost complete weighted network. The purpose of our maximum picking procedure is to enhance the signal to noise ratio in such a way to build a conceptually different network, whose links are directed and related to necessity instead of proximity.\\
In summary, the differences between the Taxonomy Network and the Product Space are i) the presence of directed links, with a clear causality meaning; ii) the reduction of the number of link from order $N^2$ to order $N$ and iii) the different normalization, which takes into account the different diversifications o countries.

\subsection*{Tests of the algorithm}
\subsubsection*{Sample matrices}
\begin{figure}
\centering\huge
\subfigure[$M_{cp}$ matrix]{
$\begin{turn}{90}
  \textrm{\!\!\!\!\!\!\!\!\!\!\!\!\! countries}
 \end{turn}
 \stackrel{\mbox{products}}{
\left(\begin{array}{ccccc}
1 & 1 & 1 & 0 & 1 \\
0 & 1 & 0 & 0 & 1 \\
1 & 1 & 0 & 0 & 1 \\
0 & 1 & 0 & 1 & 0 
\end{array}
\right)}$

}
\hspace{5mm}
\subfigure[Taxonomy Product Network]{
 \raisebox{-33mm}{
 \includegraphics{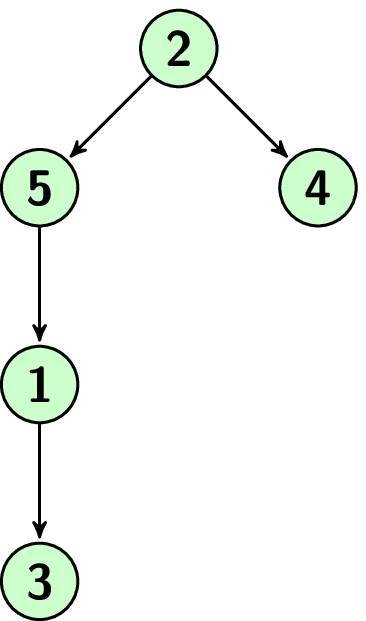}
 }

}
\caption{On the left, a sample $M_{cp}$ matrix. On the right, the resulting Taxonomy Network. One can notice how the presence of products in the export baskets of the countries influences their position in the network. For example, the ubiquitous product 2 becomes the root.}
\label{fjsjflsd}
\end{figure}

Now we give an example of the output of our algorithm, starting from a simple $M_{cp}$ matrix. We show both the matrix and the resulting taxonomy network in Fig.\ref{fjsjflsd}. Here countries are ordered in rows and products in columns; for example, the second country produces the second and the fifth product. Now we focus on the relationship between the structure of the matrix and the one of the network.

Product 2, which corresponds to the second column, is made by all the countries: this means that, probably, the capabilities needed are relatively few or simple to achieve. On the contrary, products 3 and 4 are exported by only one country, so we can argue that very specific features, which has been developed only by countries 1 and 4 respectively, are required by these products. Products 5 and 1 lay somehow in the middle. The resulting network is depicted in Fig.\ref{fjsjflsd}. The ubiquitous product 2 results to be the root, and it is needed to make all the other products. In particular, the fact that country 4 (fourth row) exports only products 2 and 4 suggests that the capabilities needed to produce 2 are a mandatory condition to produce 4. The left branch is constituted by a chain of products built following the same line of reasoning.\\

\subsubsection*{Capabilities-based model}
\begin{figure}
\centering
\includegraphics[scale=0.4]{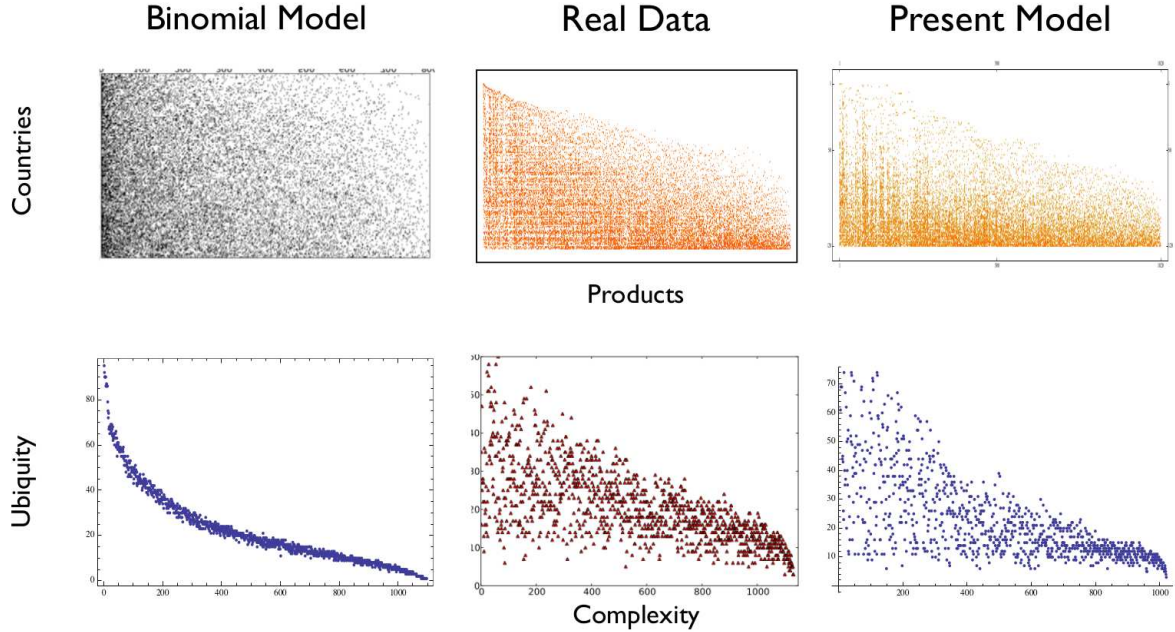}
\caption{The peculiar triangular shape of the $M_{cp}$ matrix and the empirical distribution of products in the ubiquity-complexity plane are well reproduced by our model.} \label{gskjfgjkd}
\end{figure}

In order to further test our algorithm we have built a model in which there is a well defined and known relation between the products, in the very same spirit of Fig.\ref{fhsdsljfjks}. In this way we can obtain both the taxonomy and the $M_{cp}$ matrix and we can compare the performance of various reconstruction methods.\\
\textbf{Definition of the model} The construction of the product taxonomy starts with $R$ root products. Each of these products needs only one capability in order to be produced. At this stage we intend a capability as the \textit{minimal} and \textit{non-trivial} endowments needed in order to produce a product. By \textit{non-trivial} we mean that a given capability is not owned by all the countries by default (in a real-world example a trivial capability could be water or sunlight). By \textit{minimal} we mean that a capability is the smaller set of endowments which makes the difference between being able or not to produce a new product in at least one case (in a real world example a single oil well will not make a country an oil exporter while a vast oilfield can).\\
The product taxonomy is then built as follows:
\begin{enumerate}
 \item At each time step a new capability is introduced.
 \item The new capability defines a new product $p'$ by being added at random to one of the existing products $p$ with a uniform probability.
 \item A directed link is inserted from $p$ to $p'$.
\end{enumerate}
Then the $M_{cp}$ matrix is built as follows:
\begin{enumerate}
 \item A diversification $d_c$ is assigned to each country $c$; the specific value is extracted from a real-world distribution.
 \item The country chooses randomly $d_c$ products from the taxonomy; the probability of choosing a particular product is inversely proportional to the number of capabilities (i.e. the distance from the root) associated with that product.
 \item All the products that are on the shortest path from the root of the corresponding tree to any chosen product are assigned to the country $c$.
\end{enumerate}
The values of the $d_c$ are chosen such that the distribution of the diversification in the model is similar to the one coming from the real data.\\
\textbf{Stylized facts reproduced by the model} The model is able to reproduce some non-trivial stylized facts present in the real $M_{cp}$ matrix. In Fig.\ref{gskjfgjkd} we show a comparison between a simple binomial model, in which no product taxonomy is present \cite{HH}, the real $M_{cp}$ matrix and a realization of the present model. The first row shows a representation of the matrices: it is immediately clear that the binomial model misses some key aspects, while our taxonomy based model is able to produce results much closer to the real case. This becomes clearer when looking at the scatter plot of the ubiquity vs complexity ranking of the products. Complexity is a measure of the number of needed capabilities, introduced in \cite{newmetrics,plosnm}. The peculiar triangular shape (i.e. the existence of products that are both ubiquitous and complex) present in the real-world data is very difficult to reproduce with even more complicated binomial models \cite{battistron} but emerges naturally from our 
approach. This model has 
been used to benchmark the performance of various algorithms in reconstructing the taxonomy by starting from the $M_{cp}$ matrix. In particular with 120 countries and 1120 products our algorithm is able to detect more of $80\%$ of the correct links. Full results are presented in Table \ref{hfhrthrb}.

\begin{table}[h]
 \centering
\begin{tabular}{ |l|l|l| }
\hline
\multicolumn{3}{ |c| }{\% of correctly reconstructed links} \\
\hline
 & 25 countries & 120 countries \\
 & 248 products & 1120 countries \\ \hline
Present Criteria & 42.4\% & 81.1\% \\ \hline
Max. Spanning Tree & 25.8\% & 33.4\% \\ \hline
Random links & 10.3\% & 11.3\% \\ \hline
\end{tabular}
\caption{Comparison among three different ways to reconstruct a taxonomy network. The present algorithm outperforms not only a random assignment of link, but also the maximum spanning tree obtained from the same matrix.} \label{hfhrthrb}

\end{table}

\section*{Results}
\subsection*{Analysis of the taxonomy network}
\begin{figure}
\centering
\includegraphics[scale=1]{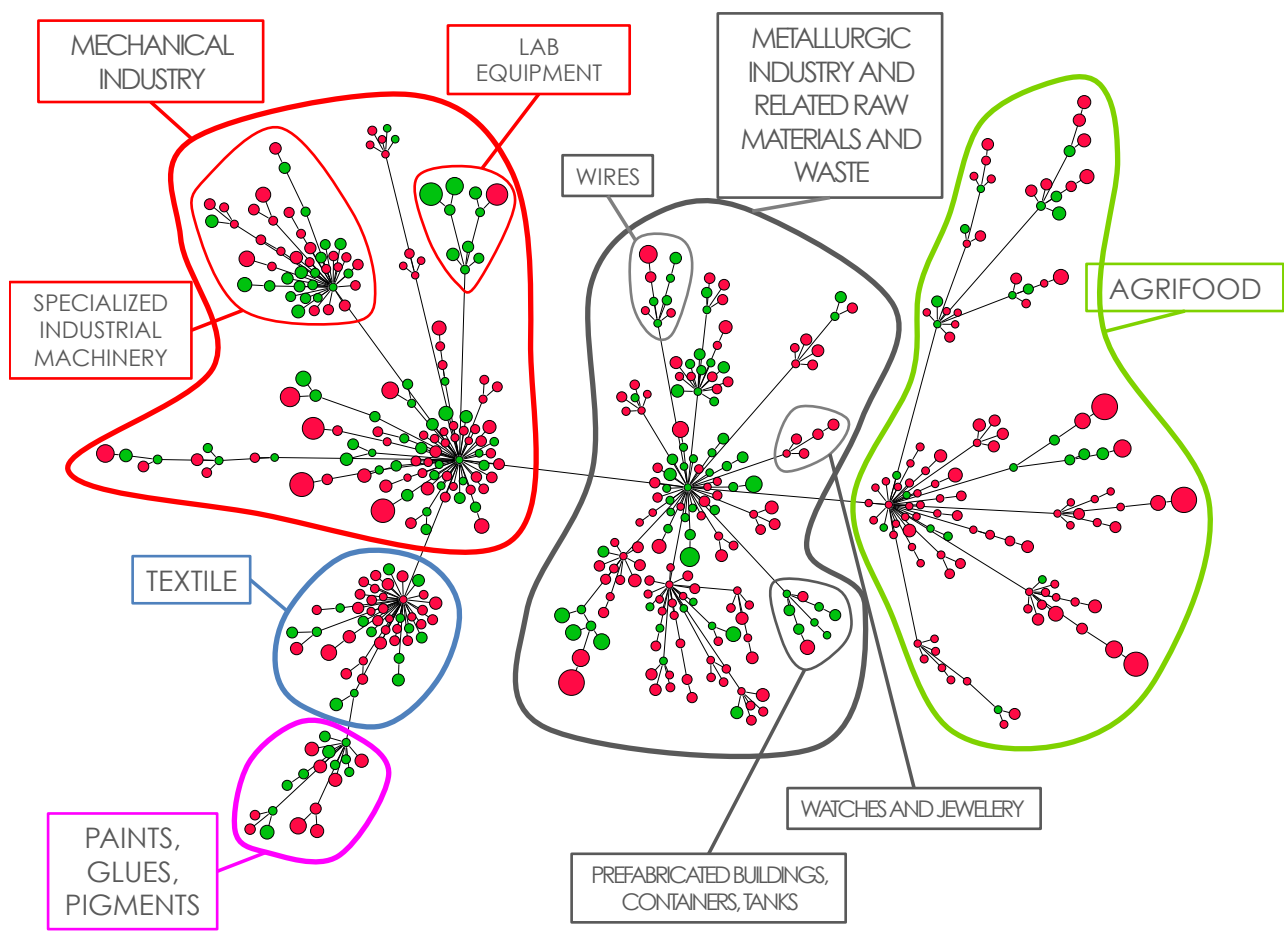}
\caption{The largest component of the taxonomy network built from the 1995-2010 database. The colors refer to the value of the $M_{cp}$ matrix for Sweden, year 2010: green is 1, red is 0. The diameter of the vertices is proportional to the logarithm of the product complexity, as defined in \cite{plosnm}.  Already from a visual inspection one could argue that a good strategic move for Sweden could be to produce the red, high complexity product in the Lab Equipment community.} \label{fksdljisd}
\end{figure}

\begin{figure}
\centering
\includegraphics[scale=0.90]{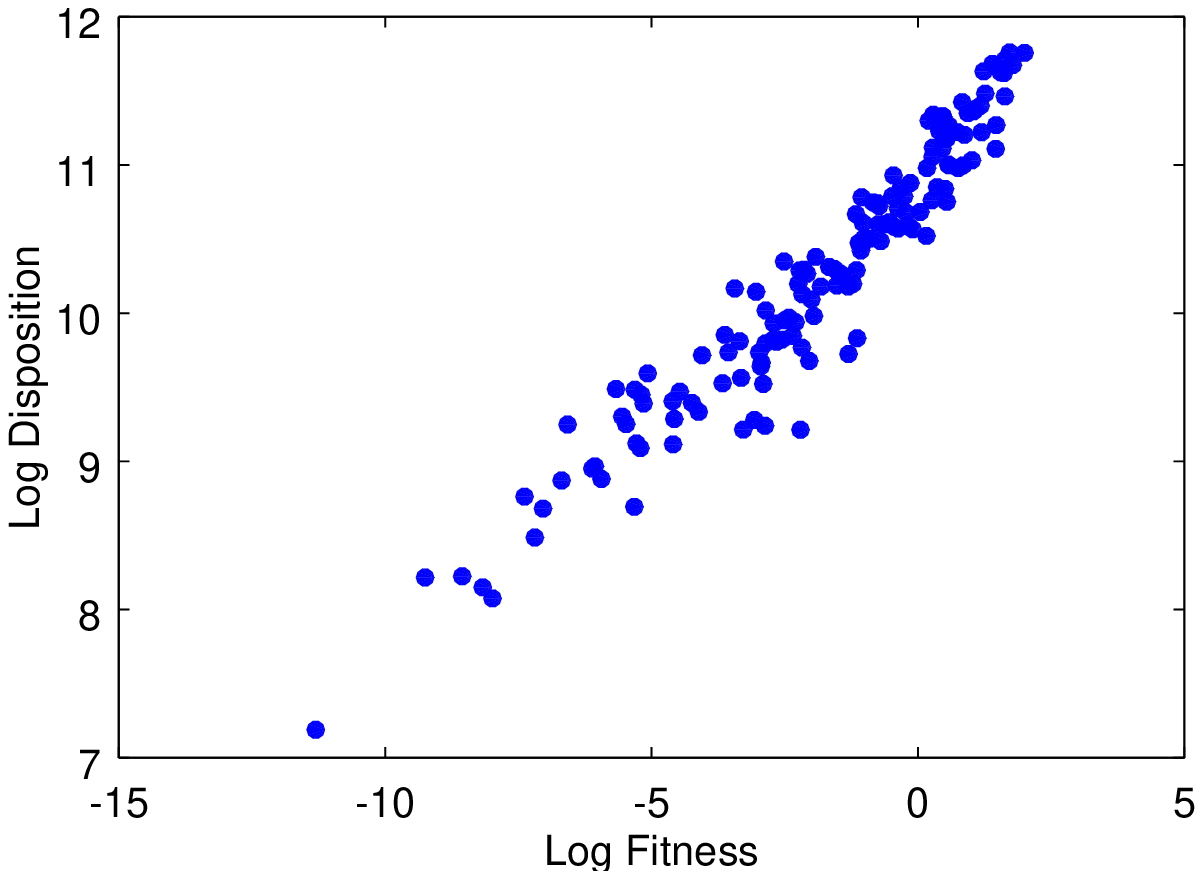}
\caption{The disposition and the fitness for each country. There is a clear correlation between the two variables, indicating a link between the growth potential of a country and its disposition on the taxonomy network.} \label{sdfwpplp}
\end{figure}

In this section we present a study of the two taxonomy networks built starting from empirical data.\\
The network we obtain from the 1995-2000 data has 1131 vertices (this number is, obviously, equal to the number of products) and 985 edges, while the 1963-2000 network has 538 vertices and 456 edges. So they are quite sparse and not fully connected (this is due to both the intrinsic heterogeneity of the products and to our filtering procedure, which selects at most one link per row. As we will see in the following, this filtering permits to identify the most relevant links from the point of view of the observed time evolution). In both networks we have about one hundred components with heterogeneous sizes. However, most of these components have a well defined economical and technological meaning. In Fig.\ref{fksdljisd} we show the largest component of the taxonomy network built from the 1995-2010 export matrices. Green filled nodes represent products that are exported by Sweden in the year 2010, while the red ones have $M_{cp}=0$. The diameter of the vertices is proportional to the logarithm of the product 
complexity, whose measure has been defined 
in \cite{newmetrics,plosnm}. One can notice a clear tendency to have products of large complexity on the border of the network, while more basic products lay in the center and have a higher degree, that is, centrality tends to be anticorrelated with complexity. This behavior is in agreement with our hypothesis that the few capabilities needed to produce \textit{low} complexity products represent a necessary condition to be able to produce \textit{high} complexity products, in the spirit of the Taxonomy Network concept we introduced in the previous sections. A zero-order validation of this idea can be found in the fact that for both networks about the 70$\%$ of the edges point from a high to a low complexity product. In a purely random framework we would expect this value to approach one half, given the presence of hundreds of edges. On the contrary, we observe a situation with a negligible p-value, and so we can conclude that the direction of links is not given by a fair coin flipping.\\
Now we want to turn our attention to how countries occupy the Taxonomy Network. In particular, we would like to study the possibility to link macroeconomic features of the countries with the properties of the vertices corresponding to the products they export. We have noticed that developed countries tend to occupy outlying vertices. In order to better study this feature we need a measure of the centrality of a given vertex which takes into account not only its degree but also the direction of the links, in such a way to pass the received authority following the links. One possible measure is the PageRank \cite{page1999pagerank}. In order to evaluate the degree of development of a given country we count its products weighting more the ones that lie away from the center, that is, vertices with a low PageRank. We study the sum of the inverse PageRank of the exported products of a given country $c$:
\begin{equation}
 D_c=\sum_{p} M_{cp} PR_p^{-1}
\end{equation} 
which we call \textit{disposition} of the country. In Fig.\ref{sdfwpplp} we plot this quantity versus the so-called fitness \cite{newmetrics,plosnm}, that is a measure of the growth potential of a country, for all the countries in our database, referring to the year 2000, finding an impressive correlation between the two ($R^2=0.92$). For clarity purposes we have taken the logarithm of both variables. This is an interesting link between a network based quantity and the fitness, which is the result of an algorithmic interplay between the countries and the complexity of the products they export.
\subsubsection*{Study of countries' development}
\begin{figure}
\centering
\includegraphics[scale=0.04]{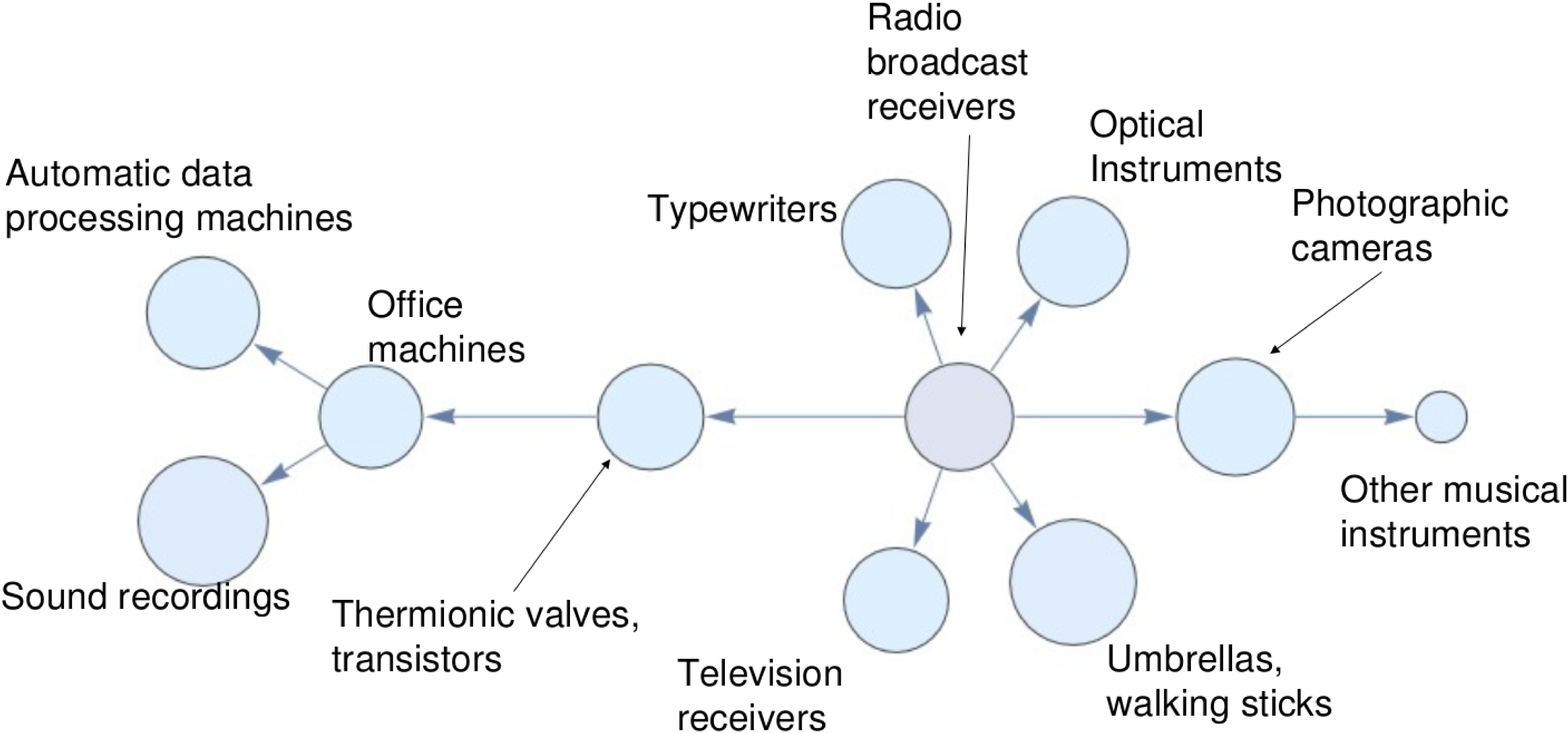}
\caption{A component of the Taxonomy Network. All nodes are clearly member of the same technological community, but \textit{umbrellas}, whose presence is due to noise.} \label{kjfshdffs}
\end{figure}

\begin{figure}
\centering
\includegraphics[scale=0.04]{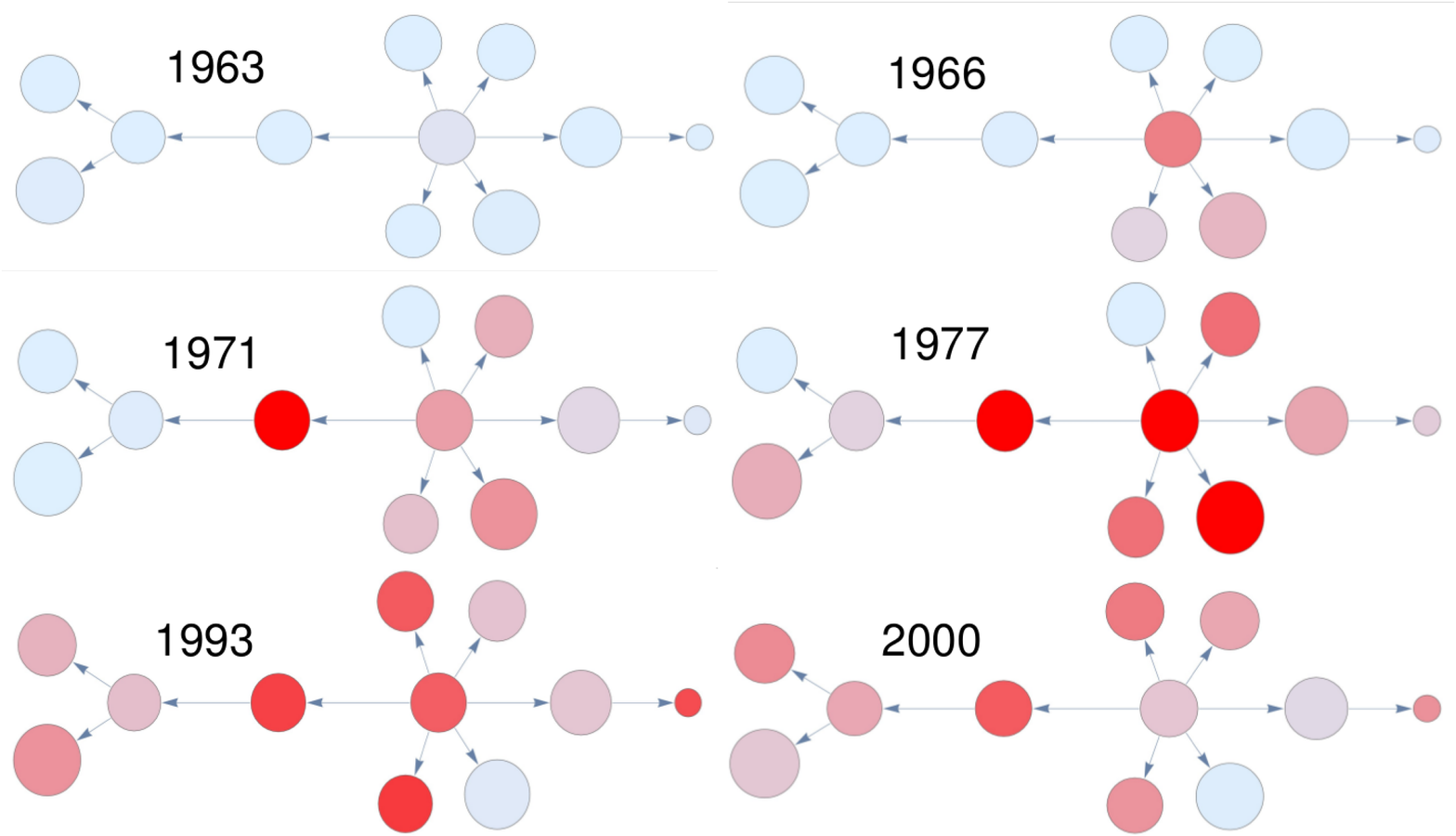}
\caption{An example of the time evolution of a component of the Taxonomy Network. The studied country is South Korea. The red fillings represent an Increase of the RCA value. One can notice the diffusion from the center (root product) towards the borders of the component.} \label{kjfshdf}
\end{figure}

\begin{figure}
\centering
\includegraphics[scale=0.8]{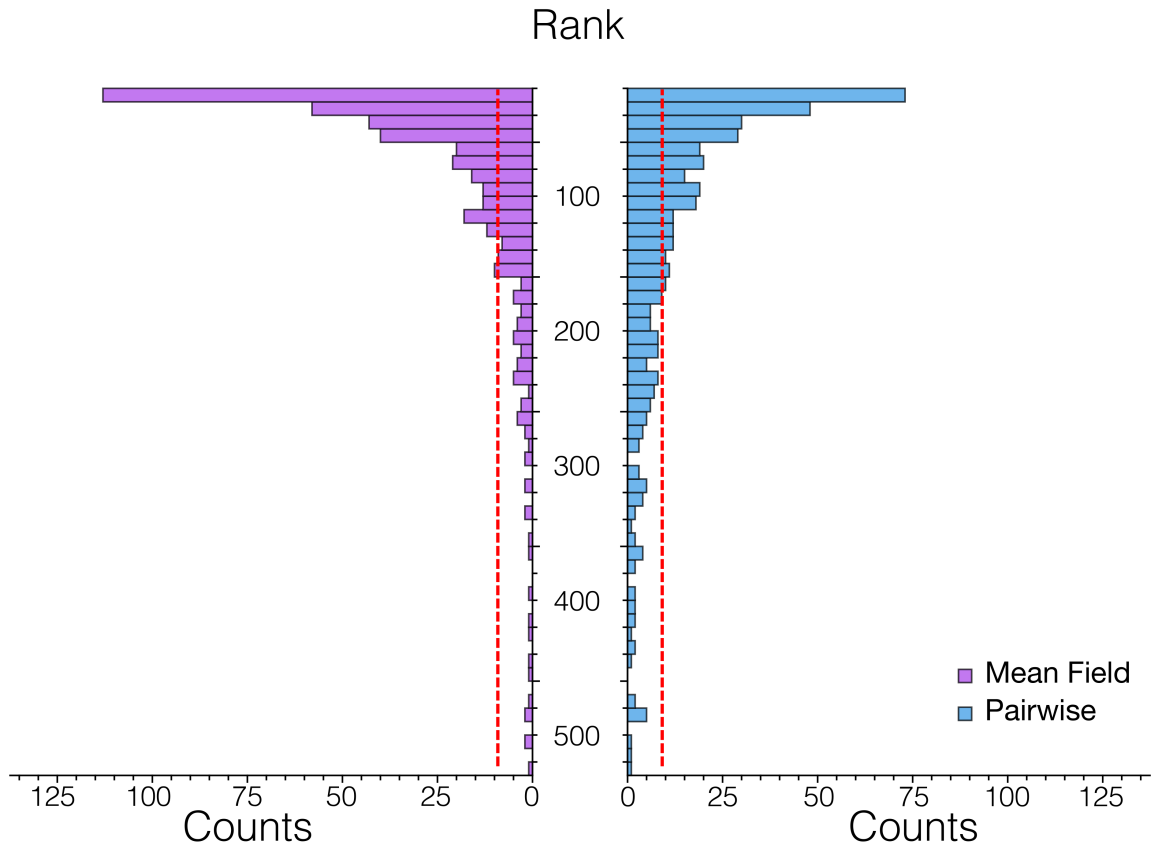}
\caption{The Taxonomy Network correctly finds the empirically most active edges, as empirically calculated with the Enabling Matrix, using two different approaches which are described in the text. Here we show the distribution of the rankings of the Enabling Matrix elements selected by the Taxonomy Network. The rankings are calculated ordering the rows of the Enabling Matrix from the largest element to the smallest one. The resulting distribution is peaked around small values, implying that a large fraction of the links suggested by our network correspond to the largest elements of the Enabling Matrices.} \label{fsfsrhiii}
\end{figure}

One of the most important features of this approach is the visualization of countries' economic development. In order to show how clearly patterns emerge when studying specific countries through time, we focus on a specific example of the development of one of the so-called ``Asian Tigers'', South Korea, which is often reported as a case study for a successful industrialization process. In particular, in Fig.\ref{kjfshdffs} we show a technological component of the taxonomy network. The root product is \textit{radio broadcast receivers}, while on the border we find \textit{automatic data processing machines}, that is, computers. An evident exception is \textit{umbrellas}, a product which obviously has nothing in common with the others and remains connected to this component despite the filtering procedure which, on the contrary, seems to perform well for the other vertices. In Fig.\ref{kjfshdf} we show the time evolution of the South Korean export for this component. The colors are proportional to Balassa's 
Revealed Comparative 
Advantage 
(RCA) \cite{balassa1965trade}: light blue means that the product is not exported, while the different shades of red are proportional to an RCA increase. In 1963 this country did not export any product of this component in a significant way. After three years, the root starts to be produced together two close products. In the following years South Korea explores the network, reaching in 1993 an impressive level of diversification. In 2000 South Korea focused its exports on borderline products, as expected from an already developed country from the disposition analysis we presented above.  The presence of a meddlesome product (in this case, \textit{umbrellas}) is due to noise, but it can be spotted thanks to its RCA behavior, which is uncorrelated with the other nodes. So, even if the probabilistic approach we use to define the network can lead to spurious results, like the presence of unexpected products in otherwise well defined clusters, one can see that a careful analysis of the dynamics clearly points to 
the fact that this site is anomalous with respect to the cluster considered.
Now we would like to study the utility of our network structure in the study of countries' development, in order to quantify on a larger scale what we observed in the South Korean test case. To do so we focus on the longest database (1963-2000) and we try to extract from the export matrices the empirical information regarding the correlation between the presence of a product in a country's export basket and the appearance of a new one in a future year. In order to do this we try to quantify how much the presence of a product $p$ influences the possible turning on of another product $p'$. One possible measure of this helpfulness is the frequency of the activations given the presence of an already activated product. In practice, first one has to calculate the three dimensional Activation Matrix
\begin{equation}
 Z_{cpy}= M_{cpy}-M_{cp(y-1)}
\end{equation}
where $y\in[1964,2000]$. For the moment we focus only on the activations of products ($ Z_{cpy}=1$) and so we ignore the cases in which $ Z_{cpy}=-1$, which corresponds to the dismission of a certain production. This different question will be addressed in a future work. In order to evaluate the frequency of activation of $p'$ given the presence of another product $p$, we calculate the Enabling Matrix 
\begin{equation}
 C_{pp'}= \frac{\sum_{c,y} Z_{cp'y} M_{cp(y-1)}}{\sum_{c,y} Z_{cp'y}}
\end{equation}
where, in the previous formulas, the matrix operations are intended as element by element operations. The elements of this matrix represent an empirical proxy of the strength of the directed link from $p$ to $p'$. Obviously, this could be a rough approximation, because in principle one can think that the more a product is present, the more it will appear to be necessary even if it could be not. For this reason we checked the weight of products' ubiquity, finding that, even if ubiquitous products tend to be more necessary, once that this effect is removed our results are substantially left unchanged. Another possibility is that it is the presence of a set of products that changes the probability that a country has to produce a new product, and not only one as supposed above. In this case it is not straightforward to calculate the relative usefulness of the products so, as a zero approximation, one can give for every activation a 
score $1/n$ to each product which was 
already exported during the previous year, where $n$ is the number of the products exported by the studied country, and so, in general, a function of $p$, $c$ and $y$. Using this approach the empirical strength of the link from $p$ to $p'$ will be given by the sum of the scores collected by the different countries through the years. In this way, the Enabling Matrix is calculated supposing a mean field interaction, in contrast with the previous approach, in which the interaction was assumed to be pairwise.\\
Once that we defined an empirical benchmark regarding the time evolution of countries' export, we have to assess its connection with the taxonomy network. To do so we sort the rows of the two enabling matrices from the largest to the smallest element and we check the position of the matrix element that we would have picked following the taxonomy network. In other words, we check how strong is the empirical realization of link present in the Taxonomy Network with respect to the other possible links. The results are depicted in Fig.\ref{fsfsrhiii}. The taxonomy network correctly identifies most of the top empirical temporal connections between products, both in the pairwise and in the mean field approach. In particular, about one hundred of the links are in the top $2.5\%$ and about an half of them are in the top $10\%$. The taxonomy network performs slightly better in the mean field case.\\
This result points out a clear connection between the taxonomy network, which is built up without considering the time evolution of the exports, and the properties of countries' development in terms of the temporal connections among the products they are exporting and the ones they will export.

\section*{Discussion}
In the present work we introduce an algorithm which is able to extract the relevant information from the time evolution of a bipartite network. In particular, we build a directed network whose nodes are constituted by only one typology of the nodes of the starting bipartite network and whose edges point from a required node to a supported node, in the sense that the activation of the first node increases the probability that the second node will be activated in the future. Having this causality relationship in mind, we named this a \textit{taxonomy network}. The algorithm, based on picking the maximums of the projection of the bipartite network, is tested on simple matrices and with a toy model which is able to reproduce the main stylized facts of the export matrices. We use this framework to analyze the taxonomy network resulting from the export data of countries. The network properties are linked to countries' potential growth and development. In particular, this last aspect is investigated by introducing 
the \textit{enabling matrix}, whose 
elements are a measure of how necessary is an activated  product to be able to produce another product in the future. We find that the largest temporal connections which we find from an empirical analysis of the export matrices are the ones we would have picked by looking at the structure of the taxonomy network.\\ This fact links the static properties of the taxonomy network with the time evolution of the bipartite one.\\
This work opens up the possibility to a number of possible applications. In general, such an algorithm could be used in any bipartite system, especially in the cases in which one topology of the nodes play an active role in choosing which node pick from the other topology, for example in the country-product networks, user-item, consumer-purchased product, and in all other recommendation systems. Regarding the country-product network, the next step would be link prediction, i.e. to build a framework able to predict which product will be exported by a given country in the next years. For example, one could look for those products which are linked to the ones which are already exported by the country in analysis: in our framework, one could say that fewer capabilities are needed to make that step. As a consequence, the taxonomy network can be also used to give policy suggestions, because a product which is close to many already produced is easier to produce. In particular, the correspondence between 
our network structure and the empirical time connections measured by the Enabling Matrix suggests that there is a well defined path to follow in the industrialization process: in the product space the possible trajectories are many, but a number of them are the preferred ones to achieve countries' growth. To simply copy other countries without learning their capabilities can not give long lasting results in terms of enduring economic stability. A less developed country has to learn simple capabilities and to be consequently able to export which we are calling the root products in order to start a stable industrialization and development process.


\section*{Acknowledgments}
We would like to thank Fabio Saracco for the essential data sanitation procedure and Emanuele Pugliese for the useful comments and discussions.

\bibliographystyle{unsrt}
\bibliography{biblio}

\end{document}